\documentstyle[11pt,aaspp4]{article}
\slugcomment{Submitted for the special HST ERO issue of the
Astrophysical Journal Letters}
\lefthead{JENKINS, ET AL.}
\righthead{HIGH VELOCITY GAS IN THE VELA SNR}
\begin{document}
\title{UV Absorption Lines from High-Velocity Gas in the Vela Supernova
Remnant: New Insights from STIS Echelle Observations of
HD72089\altaffilmark{1}}
\altaffiltext{1}{Based on observations with the NASA/ESA {\it Hubble
Space Telescope} obtained at the Space Telescope Science Institute,
which is operated by AURA, Inc. under NASA contract NAS5$-$26555. The
analysis was supported by NASA Grant NAG5-30110 to Princeton
University.}
\author{Edward~B.~Jenkins\altaffilmark{2},
Todd~M.~Tripp\altaffilmark{2}, Edward~L.~Fitzpatrick\altaffilmark{2},
Don~Lindler\altaffilmark{3,4}, Anthony~C.~Danks\altaffilmark{5,4},
Terrence~L.~Beck\altaffilmark{3,4}, Charles~W.~Bowers\altaffilmark{4},
Charles~L.~Joseph\altaffilmark{6},
Mary~Elizabeth~Kaiser\altaffilmark{7,4},
Randy~A.~Kimble\altaffilmark{4}, Stephen~B.~Kraemer\altaffilmark{8,4},
Richard~D.~Robinson\altaffilmark{9,4},
J.~Gethyn~Timothy\altaffilmark{10},  Jeff~A.~Valenti\altaffilmark{11},
Bruce~E.~Woodgate\altaffilmark{4}}
\altaffiltext{2}{Princeton University Observatory, Princeton, NJ 08544}
\altaffiltext{3}{Adv. Comp. Concepts}
\altaffiltext{4}{Code 681, NASA Goddard Space Flight Center, Greenbelt,
MD 20771}
\altaffiltext{5}{Hughes/STX}
\altaffiltext{6}{Physics \& Astronomy, P.O. Box 849, Rutgers University,
Piscataway, NJ 08855}
\altaffiltext{7}{Dept. of Physics and Astronomy, Johns Hopkins
University}
\altaffiltext{8}{Dept. of Physics, Catholic University}
\altaffiltext{9}{CSC}
\altaffiltext{10}{CRESS, York University, North York, Ontario M3J 1P3,
Canada}
\altaffiltext{11}{JILA, Campus Box 440, University of Colorado, Boulder,
CO 80309$-$0440}
\begin{abstract}
The star HD72089 is located behind the Vela supernova remnant and shows
a complex array of high and low velocity interstellar absorption
features arising from shocked clouds.  A spectrum of this star was
recorded over the wavelength range 1196.4~\AA\ to 1397.2~\AA\ at a
resolving power $\lambda/\Delta\lambda =110,000$ and signal-to-noise
ratio of 32 by STIS on the Hubble Space Telescope.  We have identified 7
narrow components of C~I and have measured their relative populations in
excited fine-structure levels.  Broader features at heliocentric
velocities ranging from $-$70 to $+130~{\rm km~s}^{-1}$ are seen in
C~II, N~I, O~I, Si~II, S~II and Ni~II.  In the high-velocity components,
the unusually low abundances of N~I and O~I, relative to S~II and Si~II,
suggest that these elements may be preferentially ionized to higher
stages by radiation from hot gas immediately behind the shock fronts.
\end{abstract}

\section{Introduction}\label{intro}

The large number of bright, early-type stars within or behind the Vela
supernova remnant (SNR) makes this remnant especially well suited for
the study of high velocity absorption features from shocked gases.   An
especially interesting example is the star HD72089, discovered by
Jenkins \& Wallerstein \markcite{1352} (1984) to have at least six
distinct velocity components in Ca~II.  HST observations of HD72089
using the G160M grating on GHRS ($\lambda/\Delta\lambda=20,000$)
revealed the presence of high velocity C~I features with extraordinarily
strong absorption coming from excited fine-structure levels, indicating
the gas is strongly compressed \markcite{2828} (Jenkins \& Wallerstein
1995, hereafter JW95).  In their study of gas-phase element abundances
in the halo of the Galaxy, Jenkins \& Wallerstein \markcite{3197} (1996,
hereafter JW96) used the pattern of element abundances in the high
velocity components toward HD72089 as a comparison standard for reduced
depletions caused by the destruction of grains in shocks.  With new
observations of this star covering the spectral region from 1196.4~\AA\
to 1397.2~\AA\ by STIS on HST, we are able to arrive at more accurate
results for the nature of the high velocity gas.

\section{Data}\label{data}

HD72089 is a star with $V=8$ and a spectral classification B5~II-III
\markcite{2510} (Houck 1978) that is about 1.7~kpc away from us, located
behind the western edge of the Vela SNR.  We present in this Letter
observations of this star taken on 30~May~1997 0022$-$0354
UT\footnote{Archive exposure index numbers O40O01P9M, PDM and PHM, with
integration times of 1420, 2560 and 2560~s, respectively.} to
demonstrate the performance of the high-resolution UV echelle
spectrograph (E140H mode) on STIS with the $0.2\times 0.09$ arc-sec
entrance aperture.

Photoevents registered by the MAMA detector were recorded in the
time-tag mode.  We created a 2048x2048 Hi-Res image (2 samples per MAMA
pixel) with corrections for the position of each time-tagged count to
compensate for (1) the orbital Doppler motion (amplitude = 10 Hi-Res
pixels) and (2) the motion caused by thermal distortions (0.2 Hi-Res
pixels per hour, measured from the slow drift of the spectral lines
within the data).  Our spectra were taken from simple, unweighted
extractions with background levels defined by the bottoms of obviously
saturated lines.  In cases where lines of interest appeared twice in
adjacent orders, the overlapping coverages were co-added.

The wavelength scale was computed using a pre-launch dispersion relation
corrected for a zero point wavelength shift computed from a wavelength
calibration observation taken with the data.  The precision of this
scale is shown by the excellent agreement for the derived wavelengths of
narrow features seen in two different orders.  Also, the velocities of
our C~I components agree to within $2~{\rm km~s}^{-1}$ (rms error) with
those of Na~I observed by Danks \& Sembach \markcite{2965} (1995).

Except in the vicinity of the strong stellar Ly-$\alpha$ line, our
derived spectrum has a signal-to-noise ratio of about 32.  From a narrow
line of Cl~I (not discussed here), we estimate that the instrumental
profile is a Gaussian with a FWHM equal to $\lambda/1.1\times 10^5$.  In
this Letter, we focus principally on the absorption lines of C~I
(\S\ref{CI}) with their noteworthy levels of fine-structure excitation,
and in \S\ref{atoms} we report on some key elements that show up in
ionization stages that are favored in normal H~I regions.

\section{High-Excitation C I}\label{CI}

In the ISM, the excited $^{3}P_{1}$ and $^{3}P_{2}$ fine-structure levels (denoted
C~I$^*$ and C~I$^{**}$) in 
the ground state of \ion{C}{1} 
can be populated by collisions with neutral hydrogen atoms, electrons and protons
\markcite{1847, 1034, 2214, 3409} (Bahcall \& Wolf 1968; Jenkins \& Shaya 1979;
Keenan 1989; Roueff \& Le Bourlot 1990).  In the 
previous {\it HST} study of HD 72089, \markcite{2828}JW95 measured remarkably
strong \ion{C}{1}$^*$ and \ion{C}{1}$^{**}$ lines in the 
component at $v=+121~{\rm km~s}^{-1}$.  The large implied gas 
pressure suggested that this \ion{C}{1} absorption arises in compressed gas 
following a shock inside a cloud that had been hit by the supernova blast wave. 
Our new STIS observations of 
HD 72089 have substantially improved resolution and provide an opportunity to 
verify and refine the results of \markcite{2828}JW95.

\begin{figure}
\plotone{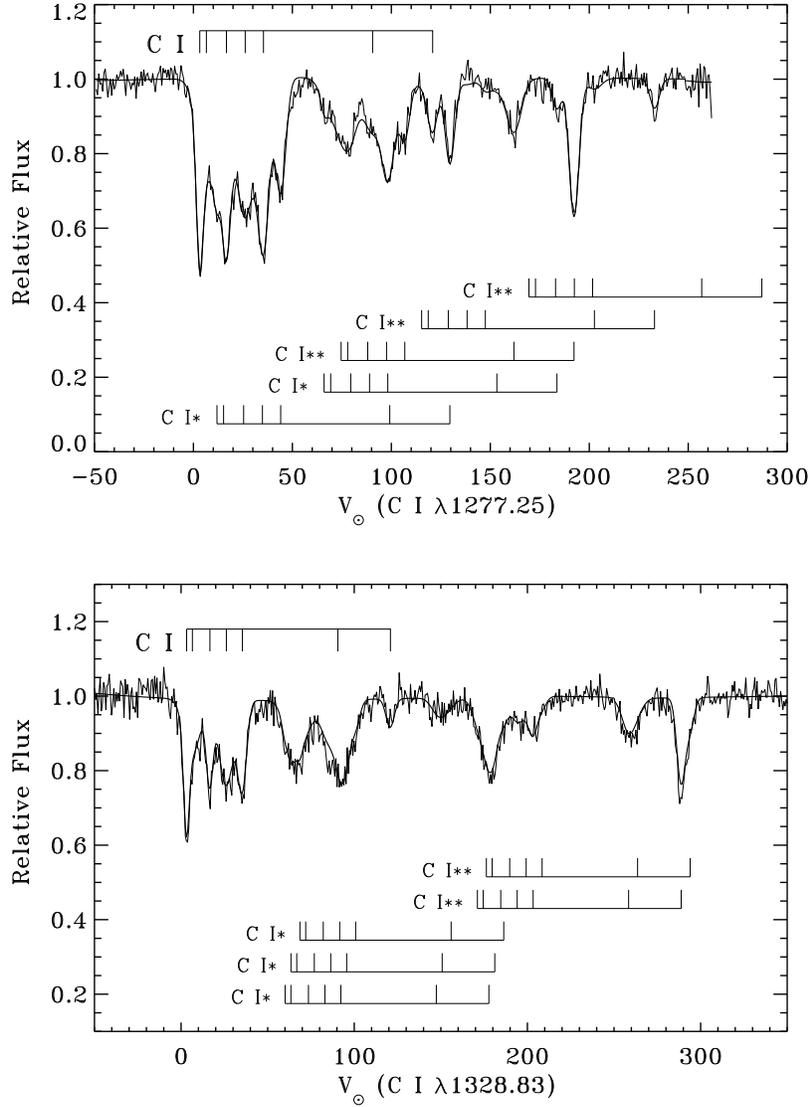}
\vspace{-2cm}
\caption[]{Absorption profiles in the heliocentric velocity frame of the (top) 
C~I 1277 \AA\ and (bottom) C~I 1329 \AA\ multiplets. The profile fits based on
the component parameters in Table \protect~\ref{cextab} are overplotted 
on the data, and the velocities of the seven components in the various C~I, 
C~I$^*$, and C~I$^{**}$ lines in these multiplets are indicated. The C~I multiplet 
near 1280 \AA\ was also fitted but is not shown here.\label{cifig}}
\end{figure}

We analyzed \ion{C}{1} multiplets
near 1277 \AA\/, 1280 \AA\/, and 1329 \AA\/ using a component 
fitting technique \markcite{2462, 3329} (Spitzer \& Fitzpatrick 1993; Fitzpatrick
\& Spitzer 1997) that minimizes the $\chi^2$ between the fit and the data to
determine 
the set of velocity centroids, $b$-values, and column densities, assuming the
original intensities were convolved with an instrumental line spread function
consisting of a Gaussian with FWHM = 4 Hi-Res pixels.  A comparison of the derived
fits with the recorded spectra near the 1277~\AA\/ and
1329~\AA\/ \ion{C}{1} absorption complexes is shown in Figure ~\ref{cifig}. The 
parameters for the 7 individual absorption 
components are given in Table~\ref{cextab}.  The associated 1$\sigma$ errors
include the effects of random photon noise, residual fixed pattern 
noise and uncertainties in determining the continuum.
Since we relied on empirically derived backgrounds, there
may be additional systematic errors at the $\sim$5\% level.
Such errors, however, will tend to shift all of the \ion{C}{1}, \ion{C}{1}$^*$,
and \ion{C}{1}$^{**}$ 
column densities in a given multiplet in the same direction and thus will not
significantly affect the derived population ratios, $f_1$ and $f_2$, listed in the
last two columns of the table.  Figure ~\ref{cifig} indicates 
that many lines are free from blending, and consequently the component 
parameters are well-constrained by the data.

\placetable{cextab}
\begin{deluxetable}{lccccccc}
\tablewidth{0pc}
\tablecaption{Profile Fitting Results: Component Velocities, Doppler 
Parameters, and Column Densities for \ion{C}{1}, C~I$^*$, and
C~I$^{**}$\label{cextab}}
\tablehead{Component & $v_{\odot}$\tablenotemark{a} &
$b\tablenotemark{b}$ & log $N$(\ion{C}{1}) & log $N$(\ion{C}{1}$^*$) &
log $N$(\ion{C}{1}$^{**}$) & $f_{1}\tablenotemark{c}$ &
$f_{2}\tablenotemark{c}$ \nl
Number & (km s$^{-1}$) & (km s$^{-1}$) & \ & \ & \ & \ & \nl}
\startdata
1..... & 3.2$\pm$0.1 & 1.6$\pm$0.3 & 12.84$\pm$0.02 & 12.29$\pm$0.08 &
$<11.7$\tablenotemark{d} & 0.22 & $<0.054$ \nl
2..... & 6.5$\pm$0.2 & 6.3$\pm$0.7 & 12.96$\pm$0.02 & 13.06$\pm$0.02 &
12.92$\pm$0.02 & 0.39 & 0.29 \nl
3..... & 16.7$\pm$0.1 & 2.2$\pm$0.4 & 12.78$\pm$0.02 & 11.54$\pm$0.41 &
$<11.6$\tablenotemark{d} & 0.05 & $<0.059$ \nl
4..... & 26.2$\pm$0.1 & 5.3$\pm$0.6 & 13.11$\pm$0.02 & 12.85$\pm$0.03 &
12.60$\pm$0.04 & 0.29 & 0.17 \nl
5..... & 35.4$\pm$0.1 & 3.2$\pm$0.2 & 12.94$\pm$0.02 & 12.97$\pm$0.01 &
12.59$\pm$0.03 & 0.43 & 0.18 \nl
6..... & 90.5$\pm$0.2 & 5.0$\pm$0.7 & 12.38$\pm$0.06 & 12.50$\pm$0.04 &
12.72$\pm$0.02 & 0.29 & 0.49 \nl
7..... & 120.9$\pm$0.1 & 2.9$\pm$0.2 & 12.36$\pm$0.03 & 12.79$\pm$0.02 &
13.00$\pm$0.01 & 0.34 & 0.54 \nl
\enddata
\tablenotetext{a}{Errors in the formal fits exclude systematic
uncertainties in the STIS wavelength scale and laboratory wavelengths. 
Subtract $16.9~{\rm km~s}^{-1}$ to obtain LSR velocities.}
\tablenotetext{b}{Errors do not include the uncertainty in the assumed
width of the instrumental line spread function.}
\tablenotetext{c}{The relative populations of the \ion{C}{1}
fine-structure 
levels in the notation of Jenkins \& Shaya\markcite{js79} (1979): $f_{1}
= N($\ion{C}{1}$^*)/N($\ion{C}{1}$)_{\rm total}$ and $f_{2} =
N($\ion{C}{1}$^{**})/N($\ion{C}{1}$)_{\rm total}$.}
\tablenotetext{d}{Formal result for the column density is a small
negative number; upper limit quoted here reflects that result plus a
$2\sigma$ error.}
\end{deluxetable}
\placefigure{cifig}

At sufficiently high densities in a warm gas, the level populations
approach their relative statistical weights, i.e., 1:3:5, so that
$f_{1} \ \rightarrow 0.33$ and $f_{2} \ \rightarrow 0.56$.  We confirm
the determination by \markcite{2828}JW95 that this condition is found
for the component at the highest velocity (our Component 7).  We find
that some components at low and intermediate velocities (Components 2, 4
and 6) have values of $f_1$ and $f_2$ that must arise from blends of
unresolved components, each with differing degrees of C~I excitation
caused by very different pressures or levels of ionization.\footnote{See
\S IVa of Jenkins \& Shaya \protect\markcite{1034} (1979) for a simple
geometrical method of interpreting possible combinations of absorbing
regions that blend together and produce measured $(f_1,f_2)$ that are
inconsistent with a single source region.}  This conclusion is supported
by the slightly larger $b$-values of these components.  Finally,
Component 3 has an unusually low excitation.  Gas within this region
would have a pressure $p/k\approx 1300~{\rm cm}^{-3}$K if $T=300$K (and
electron and proton densities are low enough to be neglected), but $p/k$
could be as large as $7400~{\rm cm}^{-3}$K at the upper limit for
temperature $T=mb^2/2k=3400$K indicated by our measurement of the line
width.

\section{Selected Atoms and Ions}\label{atoms}

\begin{figure}
\epsscale{0.6}
\plotone{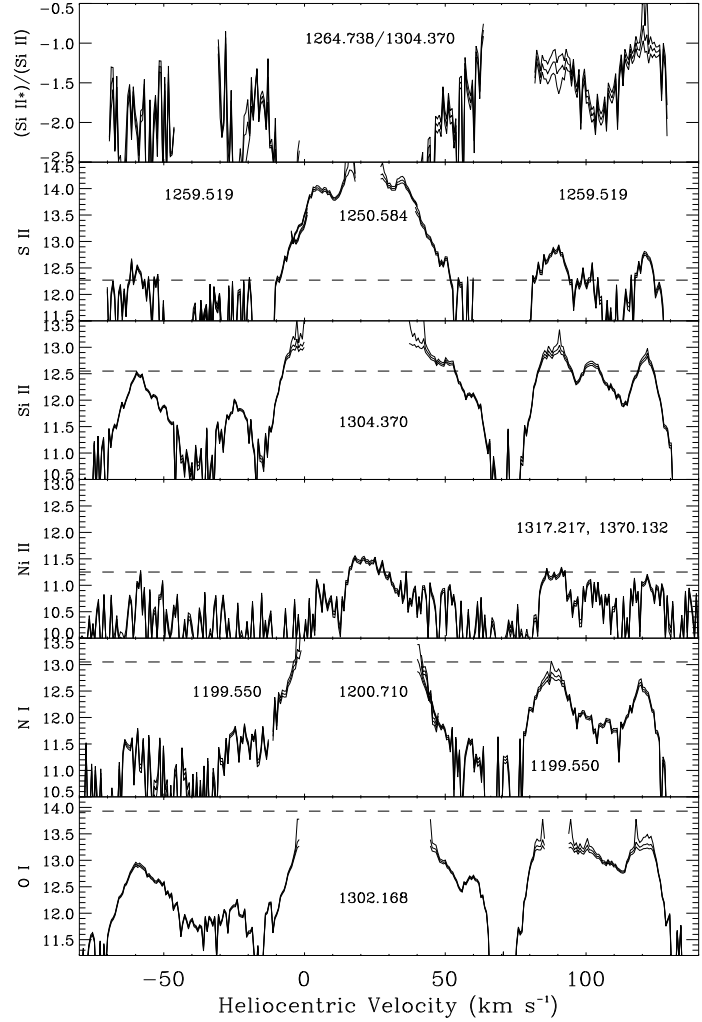}
\caption{Plots of $\log N(v)$ for 5 species, as labeled (lower panels),
and the relationship $\log [N({\rm Si~II}^*)/N({\rm Si~II})]$ (top
panel) -- an indicator of the local electron density, $n_e$ (see
Eq.~\protect\ref{n_e}).  The dashed lines (drawn at an arbitrary level
for S~II) represent levels that are consistent with the elements' cosmic
abundance relative to S, i.e., the fact that the peaks of some species
(e.g. N~I and O~I) are substantially displaced downwards from the dashed
lines relative to their counterparts for S~II indicates that these
elements are less abundant than expected, relative to S.  Transition
f-values were taken from Morton \protect\markcite{96} (1991), except for
Si~II \protect\markcite{2462} (Spitzer \& Fitzpatrick 1993) and
Si~II$^*$ \protect\markcite{2884} (Luo, Pradhan, \& Shull
1988).\label{stack}}
\end{figure}

The general character of the velocity components for ionization stages
of elements that have an IP~$>$~13.6~eV is very different from the
narrow components seen in C~I.  C~I emphasizes mostly regions that have
large internal densities, while lines from such species as C~II, N~I,
O~I, Si~II, S~II, and Ni~II are likely to arise from a much broader
range of conditions in generally lower density clouds.  It is no
surprise that the latter group display lines that have a broad velocity
extent with virtually no evidence for narrow, unresolved substructures
(see below).  For this reason, we choose to analyze the features in the
context of their apparent optical depths $\tau_a$ as a function of
radial velocity $v$ \markcite{110, 3184} (Savage \& Sembach 1991;
Jenkins 1996), a concept that is more general than component fitting. 
For the line complexes of N~I, O~I, Si~II and S~II, the saturation at
low velocities was too large to allow any column density determinations
(see Fig.~\ref{stack}).  The lines of C~II and C~II$^*$ near 1335~\AA\
were hopelessly saturated at all velocities.

To within the noise fluctuations, the high-velocity components of the
weakest lines of N~I, S~II and Si~II$^*$ yield values of apparent column
density, $N_a(v)=3.768\times 10^{14}\tau_a(v)/(f\lambda)~{\rm
cm}^{-2}({\rm km~s}^{-1})^{-1}$, that are consistent with those from the
strongest lines.  The lines for these three species cover ranges of
$f\lambda$ that differ by factors of 3, 3 and 10, respectively. 
Unfortunately, the strong line of Si~II at 1260.422~\AA\ has very
serious interference from the line of Fe~II at 1260.533~\AA, so we can
not compare it with the weaker feature at 1304.370~\AA.  The good
agreements indicate that the derived $\tau_a(v)$'s are not
underestimated because of the presence of saturated substructures that
are not resolved by the instrument (and this probably applies to many
species other than N~I, S~II and Si~II$^*$).  Thus it is not necessary
to invoke the special correction procedure outlined by Jenkins
\markcite{3184} (1996), and henceforth we will replace $N_a(v)$ with
$N(v)$, the {\it true} column density per unit velocity.

\placefigure{stack}

The lowest 5 panels in Fig.~\ref{stack} show $\log N(v)$ for S~II,
Si~II, Ni~II, N~I and O~I, with the ordering of these species being
based on an {\it apparent} decrease in general abundances below their
respective cosmic abundances \markcite{68, 3418} (Anders \& Grevesse
1989; Grevesse \& Noels 1993) (but see \S\ref{ionization}), assuming
that sulfur is undepleted \markcite{3329} (Fitzpatrick \& Spitzer 1997). 
The magnitudes of these deficiencies for the most negative and the most
positive velocity peaks are listed in the second column of
Table~\ref{deficiencies}.  For some species, there are velocity regions
where the features are either lost in the noise or badly saturated.  We
have omitted these parts of the tracings from the figure.  Absorption by
low-velocity P~II $\lambda 1301.874$ could be adding to the right-hand
side of the O~I peak centered at $-60~{\rm km~s}^{-1}$.  The plotted
lines consist of a superposition of three determinations (often
overlapping) --- a middle line that represents our best estimate for
$\log N(v)$ that is flanked by ones that represent the extremes that
could be caused by worst combinations of systematic errors ($1\sigma$)
in either the zero intensity level or the placement of the continuum
level (errors caused by noise are {\it not included}, as one can
estimate these from the size of the random wiggles in the tracings).

\placetable{deficiencies}
\begin{deluxetable}{cccc}
\tablewidth{0pc}
\tablecolumns{4}
\tablecaption{Observed and Calculated Deficiencies\label{deficiencies}}
\tablehead{
\colhead{} & \colhead{} & \multicolumn{2}{c}{$\log R$ (from
Eq.~\protect\ref{R})}\\
\cline{3-4}\\
\colhead{Species\tablenotemark{a}} & \colhead{$\log (A/{\rm S~II})_{\rm
obs} - \log (A/{\rm S})_\odot$} & \colhead{$\log T=2.7$} &
\colhead{$\log T=3.4$}\\}
\startdata
N~I ($-58$)&$-1.72$&$-2.04$\tablenotemark{b}&$-2.39$\nl
N~I (+121)&$-0.87$&$-0.93$\tablenotemark{b}&$-1.17$\nl
O~I ($-$58)&$-1.14$\tablenotemark{c}&$-1.35$&$-2.01$\nl
O~I (+121)&$-0.90$&$-0.42$&$-0.87$\nl
Al~II ($-$58)&$-0.58$\tablenotemark{d}
&$-0.07$\tablenotemark{b}&$-0.18$\nl
Al~II (+121)&$-0.66$\tablenotemark{d}&e&$-0.03$\nl
Si~II ($-$58)&$-0.36$&e&$-0.03$\nl
Si~II (+121)&$-0.21$&e&0.00\nl
S~II ($-$58)&(0.00)&$-0.12$\tablenotemark{b}&$-0.24$\nl
S~II (+121)&(0.00)&e&$-0.04$\nl
Fe~II ($-$58)&$-0.82$\tablenotemark{d}&f&$-0.75$\nl
Fe~II (+121)&$-0.35$\tablenotemark{d}&e&$-0.02$\nl
Ni~II ($-$58)&g&f&$-0.74$\nl
Ni~II (+121)&$-0.51$&e&$-0.03$\nl
\enddata
\tablenotetext{a}{Numbers in parentheses identify velocity components:
``$-58$'' applies to $\int N(v)dv$ (for column 2) over the heliocentric
velocity range $-70$ to $-45~{\rm km~s}^{-1}$ where
Eq.~\protect\ref{n_e} gives $n_e=4.6$ and $5.3~{\rm cm}^{-3}$ for $\log
T=2.7$ and 3.4, respectively, and ``+121'' applies to the range +112 to
$+130~{\rm km~s}^{-1}$ ($n_e=35$ and $39~{\rm cm}^{-3}$).}
\tablenotetext{b}{Numbers neglect charge exchange reactions with H
because the fitting formulae for cross sections are not valid for
$T<10^3$K, but judging from the small effect at $10^{3.4}$K, the
omission of these reactions is probably not important.}
\tablenotetext{c}{May include some contamination from the 1301.874\AA\
feature of P~II at low velocities.}
\tablenotetext{d}{Column density from JW96, compared to S~II measured in
this investigation.}
\tablenotetext{e}{Since $\log R>-0.1$ at $T=10^{3.4}$K, the same
probably applies here, even though we can not include charge exchange
reactions.}
\tablenotetext{f}{Charge exchange reactions have a big effect at
$T=10^{3.4}$K, thus we have no confidence in quoting a number that
neglects charge exchange at lower $T$.}
\tablenotetext{g}{Some Ni~II is present, but the absorption feature is
too weak to measure.}
\end{deluxetable}

\section{Ionization by Radiation from Shocks}\label{ionization}

The strong deficiencies of N and O shown in Table~\ref{deficiencies} are
inconsistent with previous findings for the general ISM that indicate
these two elements are only mildly depleted \markcite{14,  1966, 2050,
2858} (Hibbert, Dufton, \& Keenan 1985; Cardelli, Savage, \& Ebbets
1991; Cardelli et al. 1991; Meyer et al. 1994).  One possible
interpretation of this phenomenon is that ionizing radiation from nearby
shock fronts moves the atoms to higher, unseen stages of ionization.  To
investigate the plausibility of this idea for these two species plus
others, we performed a simple exercise.  For representative values of
$N$(Si~II$^*$)/$N$(Si~II) of $10^{-2}$ and $10^{-1.15}$ for the
components centered at $-58$ and $+121~{\rm km~s}^{-1}$, respectively,
we used the collision cross section and radiative decay rate of the
upper fine-structure level given by Keenan et al. \markcite{2389} (1985)
to obtain the electron density as a function of temperature,
\begin{equation}\label{n_e}
n_e={8.97T^{0.5}[N({\rm Si~II}^*)/N({\rm Si~II})]\over \exp (-413{\rm
K}/T) - 0.5 [N({\rm Si~II}^*)/N({\rm Si~II})]}~.
\end{equation}
We then used the $13-54$~eV ionizing radiation field of \markcite{1800}
Shull \& McKee  (1979) for a shock with a velocity $v_s=100~{\rm
km~s}^{-1}$ and preshock density $n_0=10~{\rm cm}^{-3}$ (see
\markcite{2828}JW95) and evaluated the ionization equilibrium between
the visible stages in Table~\ref{deficiencies} (denoted with a subscript
1) and the next two higher ionization states (subscripts 2 and 3) to
obtain the ratio,
\begin{equation}\label{R}
R\equiv {n_1\over n_1+n_2+n_3}=\Bigg\{ 1 + {(\Gamma_1+\delta_2n_e)
[\Gamma_2+(\alpha_2n_e+\delta_3^\prime n_{\rm H})]\over (\alpha_1n_e +
\delta_2^\prime n_{\rm H})(\alpha_2n_e+\delta_3^\prime n_{\rm
H})}\Bigg\}^{-1}~,
\end{equation}
with $n_{\rm H}=n_e^2\alpha_{n\geq 2,{\rm H}}/\Gamma_{\rm H}$. 
$\Gamma_1$ and $\Gamma_2$ are the photo-ionization rates out of the two
lower stages, using cross sections calculated from the analytic
approximations of Verner, et al. \markcite{3404} (1996) [and Verner \&
Yakovlev \markcite{3408} (1995) for Ni].  The recombination coefficients
$\alpha_1$ and $\alpha_2$ were evaluated from the parameters for the
fitting equations given by Shull \& Van Steenberg \markcite{1936} (1982)
[and Aldrovandi \& P\'equignot \markcite{1934} (1974) for Al].  Values
for the charge exchange rates $\delta_3^\prime$ (${\rm X}_3+{\rm
H}\rightarrow {\rm X}_2+{\rm H}^+$) and $\delta_2^\prime$ (${\rm
X}_2+{\rm H}\rightarrow {\rm X}_1+{\rm H}^+$) were derived from the fits
given by Kingdon \& Ferland \markcite{3407} (1996).  Unfortunately, for
all elements except oxygen, these charge exchange fits are valid only
for $T>10^3$K.  We neglected any possible self-absorption of the
ionizing radiation by neutral hydrogen, and we did not include charge
exchange with He.  At $\log T\approx 2$ only the lower of the two
Si~II$^*$/Si~II cases has an acceptable solution in Eq.~\ref{n_e}
(giving $n_e=80~{\rm cm}^{-3}$) and values of $\log R$ were about equal
to 0 for all species except for N~I, where $\log R=-0.32$.  Values of
$\log R$ at two representative higher temperatures are given in columns
3 and 4 of Table~\ref{deficiencies}.  At temperatures much above those
given in the table, collisional ionization becomes important. Post-shock
gases that are cooling radiatively through temperatures of order $10^4$K
can likewise exhibit depressions of O~I and N~I \markcite{356, 327}
(Trapero et al. 1996; Benjamin \& Shapiro 1997).

The results shown in Table~\ref{deficiencies} indicate that the apparent
deficiencies of O and N in the high-velocity components might be due to
ionization by radiation from the associated immediate post-shock gas
and/or nearby shocks.  They also indicate that measurements of Fe~II and
Ni~II in such components could, to a lesser extent, be
under-representing the true gas-phase abundances of these elements when
the electron densities are low.  Measurements of Al~II, Si~II and S~II
would still yield reliable abundances, and thus our observed 0.6~dex
deficiency of Al~II indicates that this element may still be locked up
in the remnants of the dust grains that survive the passage of a shock
(note that \markcite{3197}JW96 saw very low amounts of Al~III at high
velocity).

\end{document}